%% file: main.tex
\documentclass[10pt,conference]{IEEEtran}
\IEEEoverridecommandlockouts

\usepackage{array}
\usepackage{listings}
\usepackage{xcolor}
\usepackage{todonotes}
\usepackage{multicol}
\usepackage{subfig}
\usepackage{booktabs}
\usepackage{colortbl}
\usepackage{multirow}
\usepackage{bm}
\usepackage{tcolorbox}
\usepackage{array}
\usepackage{hyperref}
\usepackage{url}
\usepackage{threeparttable}
\usepackage{diagbox}
\usepackage{enumitem}
\usepackage{pifont}
\usepackage{microtype}
\usepackage{titlesec}
\usepackage{cite}
\usepackage{amsmath,amssymb,amsfonts}
\usepackage{algorithmic}

\newcommand{\codeff}[1]{\textcolor{blue}{\texttt{\small #1}}}

\author{
	\IEEEauthorblockN{
		Xiaoning Ren$^1$, 
		Yuhang Ye$^1$, 
		Xiongfei Wu$^2$, 
		Yueming Wu$^3$, 
		Yinxing Xue$*$ $^4$ 
	}
	\IEEEauthorblockA{
		$^1$ University of Science and Technology of China, China,
		hnurxn@mail.ustc.edu.cn, yyh834771838@mail.ustc.edu.cn \\
		$^2$ University of Luxembourg, Luxembourg, 
		xiongfei.wu@uni.lu \\
		$^3$ Huazhong University of Science and Technology, China ,
		wuyueming21@gmail.com \\
		$^4$ Institute of AI for Industries, China,
		yxxue@iaii.ac.cn
	}

}

\begin{document}

\title{Demystifying the Evolution of Neural
Networks with BOM Analysis: Insights from a Large-Scale Study of 55,997 GitHub Repositories}

\maketitle
\input{chapters/0.abstract}


\newcommand{\ie}{\textit{i.e.,} }
\newcommand{\eg}{\textit{e.g.,} }
\newcommand{\et}{\textit{et al.} }

\input{chapters/1.Introduction}

\input{chapters/2.background}
\input{chapters/3.StudyDesign}
\input{chapters/RQ1}

\input{chapters/RQ2}

\input{chapters/RQ3}
\input{chapters/5.application}
\input{chapters/threats_to_validity}
\input{chapters/8.conclusion}

\section*{Acknowledgments}
This research is supported in part by the National Defense Basic Scientific Research Program of China (grant No. JCKY2023210C009) and the National Natural Science Foundation of China (grant No. 61972373).


\bibliographystyle{IEEEtran}
\bibliography{ref}
\end{document}

%% file: chapters/0.abstract.tex
\begin{abstract}
Neural networks have become integral to many fields due to their exceptional performance. 
The open-source community has witnessed a rapid influx of neural network (NN) repositories with fast-paced iterations, making it crucial for practitioners to analyze their evolution to guide development and stay ahead of trends.
While extensive research has explored traditional software evolution using Software Bill of Materials (SBOMs), these are ill-suited for NN software, which relies on pre-defined modules and pre-trained models (PTMs) with distinct component structures and reuse patterns. Conceptual AI Bills of Materials (AIBOMs) also lack practical implementations for large-scale evolutionary analysis. To fill this gap, we introduce the Neural Network Bill of Material (NNBOM), a comprehensive dataset construct tailored for NN software. We create a large-scale NNBOM database from 55,997 curated PyTorch GitHub repositories, cataloging their TPLs, PTMs, and modules. Leveraging this database, we conduct a comprehensive empirical study of neural network software evolution across software scale, component reuse, and inter-domain dependency, providing maintainers and developers with a holistic view of its long-term trends. Building on these findings, we develop two prototype applications, \textit{Multi repository Evolution Analyzer} and \textit{Single repository Component Assessor and Recommender}, to demonstrate the practical value of our analysis.

\end{abstract}

\begin{IEEEkeywords}
Neural Network, Software Evolution Analysis, Bill of Material
\end{IEEEkeywords}



%% file: chapters/1.Introduction.tex
\section{Introduction}

Neural networks have become indispensable in modern AI due to their ability to model complex patterns and learn from large datasets.
Their usage has rapidly expanded across diverse domains, including natural language processing, computer vision, smart healthcare~\cite{SmartHealthcare}, and autonomous driving~\cite{autodrive}. This widespread adoption has led to a surge in neural network repositories, posing significant challenges for maintenance. Therefore, conducting large-scale evolution analysis and gaining a deeper understanding of the broader ecosystem's evolution are essential. Such analysis provides valuable insights into emerging trends and future directions, enabling practitioners to identify best development practices, enhance development efficiency, make informed decisions when selecting neural network technologies, and help community maintainers monitor the ecosystem's status and optimize maintenance strategies.

Existing research on software evolution has provided valuable insights into the development of traditional software systems, primarily through the analysis of software components and dependency relationships, which are often documented in a Software Bill of Materials (SBOM). 
From the perspective of software components, analyzing evolution helps track the introduction, modification, and removal of libraries, modules, and functionalities, providing insights into how systems improve in complexity and efficiency~\cite{kim2005using, juergens2009code, barbour2018investigation, thongtanunam2019will, assi2024unraveling}. Regarding dependency relationships, evolution analysis uncovers how software systems become increasingly interconnected, with modules and external libraries relying on each other~\cite{kikas2017structure, wittern2016look, decan2019empirical, yang2024harnessing}. This not only aids in optimizing the architecture but also highlights potential bottlenecks or vulnerabilities in the dependency chain, ensuring smoother maintenance and upgrades. 
However, there is a notable lack of comprehensive evolution analysis specifically targeting neural network repositories.

Neural network software differs significantly from traditional software in terms of components and dependencies. Traditional software often involves detailed logical flows, including numerous conditional and loop statements, and typically relies on a wide range of third-party libraries (TPLs) and frameworks, making its dependency management complex. The relationships between these components are deeply intertwined, requiring careful attention to compatibility, updates, and potential conflicts.
In contrast, neural networks are primarily composed of modules, such as predefined layers (e.g., convolutional or pooling layers) and pre-trained models (PTMs), which are reused across various projects. This results in simpler dependencies compared to the intricate dependency structures found in traditional software.
Consequently, existing SBOMs designed for traditional software are ill-suited for analyzing the evolution of neural network software.
Meanwhile, discussions around the AI Bill of Materials (AIBOM) have largely remained conceptual, focusing on what kinds of information should be included to promote transparency, but lacking concrete implementations or empirical validation~\cite{chan2017AIBOM, xia2023empirical, stalnaker2024boms}.

To address this gap, we define a practically applicable Bill of Materials tailored for neural networks, termed the Neural Network Bill of Materials (NNBOM). Based on this definition, we construct a large-scale dataset from the open-source community to support reliable neural network evolution analysis.
Specifically, we design a systematic process for building the NNBOM database. This process includes data collection from GitHub, extraction and analysis of key neural network components—such as third-party libraries (TPLs), pre-trained models (PTMs), and custom modules—and subsequent data processing to compile the final dataset. The result is a comprehensive database of 55,997 high-quality PyTorch repositories.

Using this database, we conduct a comprehensive analysis of neural network software evolution, revealing several key insights:
\ding{172}\textbf{Software Scale Evolution}:
The growing module count alongside stable module size suggests a prevailing convention of managing complexity through finer-grained modularization.
\ding{173}\textbf{Component Reuse Evolution}:
All components show increasing integration into larger co-usage communities, with PTMs and Modules further driving functional diversification.
\ding{174}\textbf{Inter-Domain Dependency Evolution}:
Rising cross-domain reuse reflects a shift toward general-purpose, longer-lived modules, making domain adaptability a quantifiable proxy for module quality and stability.
Finally, we develop two prototype tools: \textit{Multi-repository Evolution Analyzer}, which tracks component evolution triggered by newly added repositories over a given period; and \textit{Single-repository Component Assessor and Recommender}, which analyzes the component structure of an individual repository and recommends relevant components and similar repositories.
In summary, the main contributions of this paper are as follows:
\begin{itemize}[leftmargin=*,topsep=0pt,itemsep=0ex]
\item \textbf{The First BOM-Based Evolution Analysis of Neural Network Software}: To the best of our knowledge, this is the first work to define and construct a bill of materials tailored for neural networks (NNBOM) and to conduct an evolution analysis based on it.
\item \textbf{A Comprehensive and Open Dataset}: We build and release a large-scale dataset comprising 55,997 curated high-quality PyTorch repositories~\cite{nnbom}, offering a solid foundation for future research.
\item \textbf{Multi-perspective Evaluation}: We provide an in-depth analysis of neural network software evolution from three key perspectives: software scale, component reuse, and inter-domain dependency. This approach yields valuable insights for both researchers and practitioners.
\end{itemize}





%% file: chapters/2.background.tex
\section{Background and Related Work}

\subsection{Terminology} 
\begin{figure} [t]
    \centering
    \includegraphics[width=\linewidth]{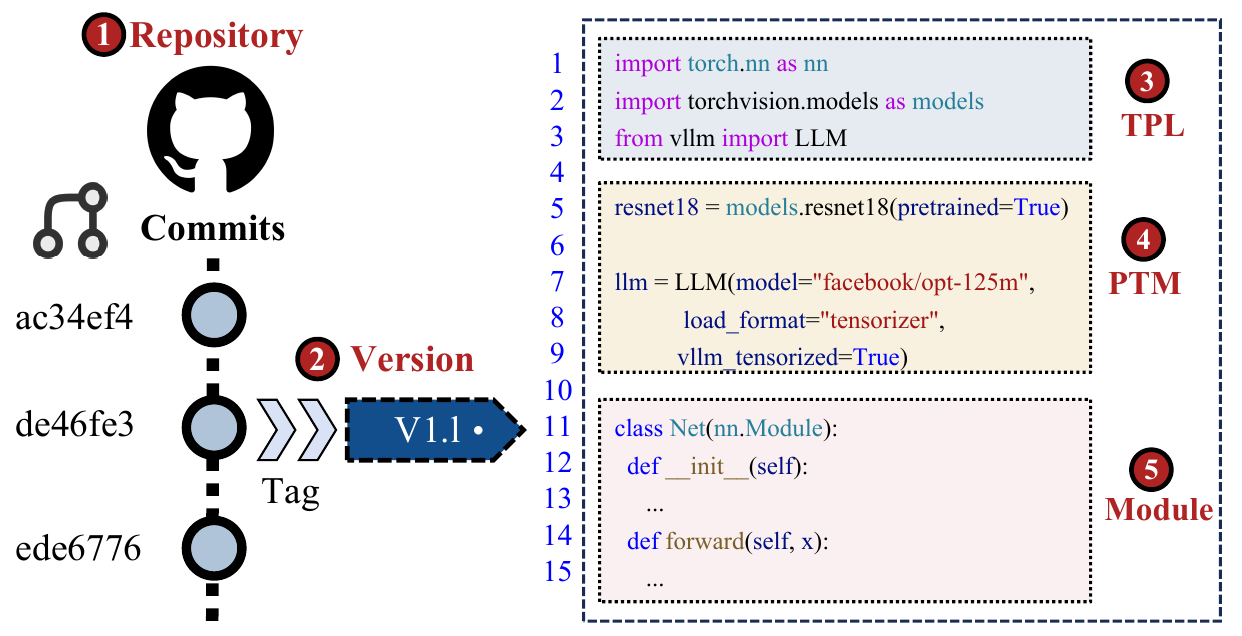}
        
    \caption{Key concepts in this research.}
    
    \label{fig: TerminologyExample}
\end{figure}

{Our study focuses on GitHub projects, primarily considering five software artifacts. Figure \ref{fig: TerminologyExample} illustrates the key concepts in this study.}

\noindent \textbf{\ding{172}Repository.} A repository is a storage location for organizing and managing project code, files, and their related version histories.

\noindent \textbf{\ding{173}Version.} A version in GitHub is a specific repository state marked by tagging a point in the commit history. These tags represent significant milestones or releases, allowing us to track and analyze the repository's development over time.

\noindent\textbf{\ding{174}Third-Party Library.} A TPL is a pre-written collection of code and resources developed by an external entity, which developers can integrate into their projects using an \codeff{import} instruction.

\noindent\textbf{\ding{175}Pre-Trained Model.} A PTM is a neural network model previously trained on a large dataset and can be used directly through function invocations.

\noindent\textbf{\ding{176}Module.} In the PyTorch framework, a module refers to any class that inherits from or indirectly derives from ``torch.nn.Module”. These modules serve as fundamental building blocks for constructing neural network architectures, encapsulating layers, and other components necessary for defining and training models.

\noindent\textbf{\ding{177}SBOM.} A Software Bill of Materials (SBOM),  typically designed for traditional software, is a comprehensive inventory detailing all libraries, frameworks, modules, and other third-party components used in software development.

\noindent\textbf{\ding{178}NNBOM.} In contrast, NNBOM is a bill of materials specifically designed to account for the unique characteristics of neural networks, such as simpler dependencies and greater code reuse, and includes components \ding{174} to \ding{176}.


\subsection{Related Work}

\noindent\textbf{Software Evolution.} 
The study of software clone evolution provides key insights into how software systems develop over time. Kim \et~\cite{kim2005using} first analyzed software clones' evolution, examining clone fragments across versions. Juergens \et~\cite{juergens2009code} conducted a large-scale case study on five software systems, revealing how different clone evolution patterns impact maintenance. Barbour \et~\cite{barbour2018investigation}, and Thongtanunam \et~\cite{thongtanunam2019will} advanced this research by constructing clone genealogies to analyze evolution patterns. ASSI \et~\cite{assi2024unraveling} extended clone evolution studies to deep learning frameworks, analyzing trends in clone code size, bug-proneness, and community size.
Beyond single projects, research has examined software evolution at a community level by analyzing inter-project clones and dependencies. Kikas \et~\cite{kikas2017structure} analyzed dependency network evolution in the JavaScript, Ruby, and Rust ecosystems. Wittern \et~\cite{wittern2016look} studied the npm ecosystem by analyzing metadata and historical changes in package dependencies, revealing the dynamic nature of dependency relationships. Decan \et~\cite{decan2019empirical} compared dependency network evolution across seven software packaging ecosystems, highlighting commonalities and differences in dependency dynamics. Yang \et~\cite{yang2024harnessing} reviewed major language models from 2018 to 2023, analyzing the rise and fall of different approaches to provide a clear understanding of large language models' development.
However, there is currently a lack of large-scale evolution analysis specifically targeting neural network software.

\noindent\textbf{SBOM.} 
SBOMs provide comprehensive transparency into the components used in traditional software systems. Carmody et al.~\cite{carmody2021medicalSBOM} demonstrated how such transparency enhances the trustworthiness, resilience, and security of medical software, benefiting software producers, consumers, and regulators.
Although various tools have been developed for SBOM generation~\cite{xia2023trust}, they are generally ineffective for neural network software due to fundamental differences, due to their model-driven architecture, and the lack of static, inspectable dependencies.
The concept of the AI Bill of Materials (AIBOM) was introduced by Chan~\et in 2017~\cite{chan2017AIBOM}. Recent work has explored how AIBOM differs from traditional SBOM~\cite{stalnaker2024boms}, and what kinds of information should be included to promote transparency and accountability in AI development~\cite{wiz2024aibom, manifest2024aibom, bennet2025implementing}. However, these discussions have largely remained theoretical, with limited practical implementation.
Given these limitations, we decided to develop NNBOM, a dataset specifically designed for neural network software, which will serve as the basis for neural network software evolution analysis.

\subsection{Motivation}

While traditional software evolution is well-studied, there exists a significant gap in understanding the unique evolutionary dynamics of NN software. To date, there has been no large-scale, component-centric evolution analysis tailored to NN systems. Existing studies often focus on traditional software, with its complex logic and dependency management, making their methods, which often rely on conventional SBOMs, less applicable to the unique characteristics of NN software. NN software is distinctly characterized by the heavy reuse of specialized components like pre-defined architectural modules and PTMs, which have different reuse patterns. While conceptual AIBOMs aim for broader AI transparency, they are not yet practical tools for systematically tracking the evolution of these core NN software artifacts at scale.

This work is motivated by the need for a practical framework to construct a Bill of Materials specifically for neural network software, enabling empirical analysis of its component-level evolution. To this end, we introduce NNBOM, a dataset that catalogs key NN software artifacts, namely third-party libraries (TPLs), pretrained models (PTMs), and custom modules. Neural network implementations are fundamentally constructed from these three types of building blocks: PTMs that are adopted for reuse, custom modules developed by practitioners, and TPLs such as PyTorch that provide the essential infrastructure. By systematically extracting these components, NNBOM holistically captures how NN software is built and how it evolves over time.
Distinct from SBOMs/AIBOMs, NNBOM's specific purpose is to enable the first large-scale, quantitative analysis of how NN software, as an ecosystem of these reusable components, evolves. Leveraging NNBOM, our study details the evolution of NN software through its scale, component reuse, and inter-domain dependencies.

%% file: chapters/3.StudyDesign.tex

\begin{figure}
    \centering
    \includegraphics[width=0.9\linewidth]{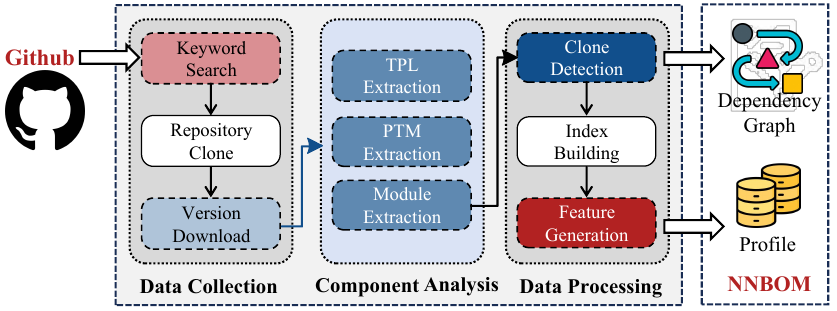}
    
    \caption{Overview of NNBOM database construction.}
    \label{fig:gene_overview}
        
\end{figure}

\section{Study Design}



This section details the methodology employed to investigate the evaluation of NN software. Our primary objective is to construct a comprehensive NNBOM dataset, capturing the constituent components of NN repositories. Building upon this dataset, we conduct a large-scale empirical analysis to understand the evolutionary patterns of NN software. The resulting NNBOM dataset is made publicly available~\cite{nnbom} to facilitate further research. The study design is structured as follows: First, we present the research questions (RQs) that guide our evolutionary analysis (Section~\ref{d1}). Second, we describe the systematic process for constructing the NNBOM database, encompassing data collection from GitHub (Section~\ref{d2}), the extraction and analysis of key NN components such as TPLs, pre-trained models (PTMs), and Modules (Section~\ref{d3}), and subsequent data processing to prepare the NNBOM for our empirical study (Section~\ref{d4}). Figure~\ref{fig:gene_overview} provides an overview of this database construction pipeline.

\subsection{Research Questions}
\label{d1}

To systematically analyze the evolution of NN software, we focus on three key facets: the changing scale of software, the reuse patterns of its core components, and the shifting landscape of inter-domain dependencies. These inquiries are formulated into the following research questions:


\begin{itemize}[leftmargin=*,topsep=0pt,itemsep=0ex] 
\item RQ1 (Software Scale Evolution): {How has the scale of NN software, particularly in Pytorch-based repositories, evolved in open-source communities, considering the number of versions, TPLs, PTMs, modules, and lines of code?} 
\item RQ2 (Component Reuse Evolution): {How have component co-usage patterns evolved in neural network software, and what structural trends can be observed in the long-term evolution of core Modules?} 
\item RQ3 (Inter-Domain Dependency Evolution): {How have inter-domain dependencies, evidenced by module reuse across different AI application domains, evolved within the NN software ecosystem?} 
\end{itemize}

The methodologies for addressing these RQs, along with their corresponding results and findings, are detailed in Sections~\ref{sec:RQ1}, \ref{sec:RQ2}, and \ref{sec:RQ3}, respectively.




\begin{figure}[t]
    \centering
    \includegraphics[width=\linewidth]{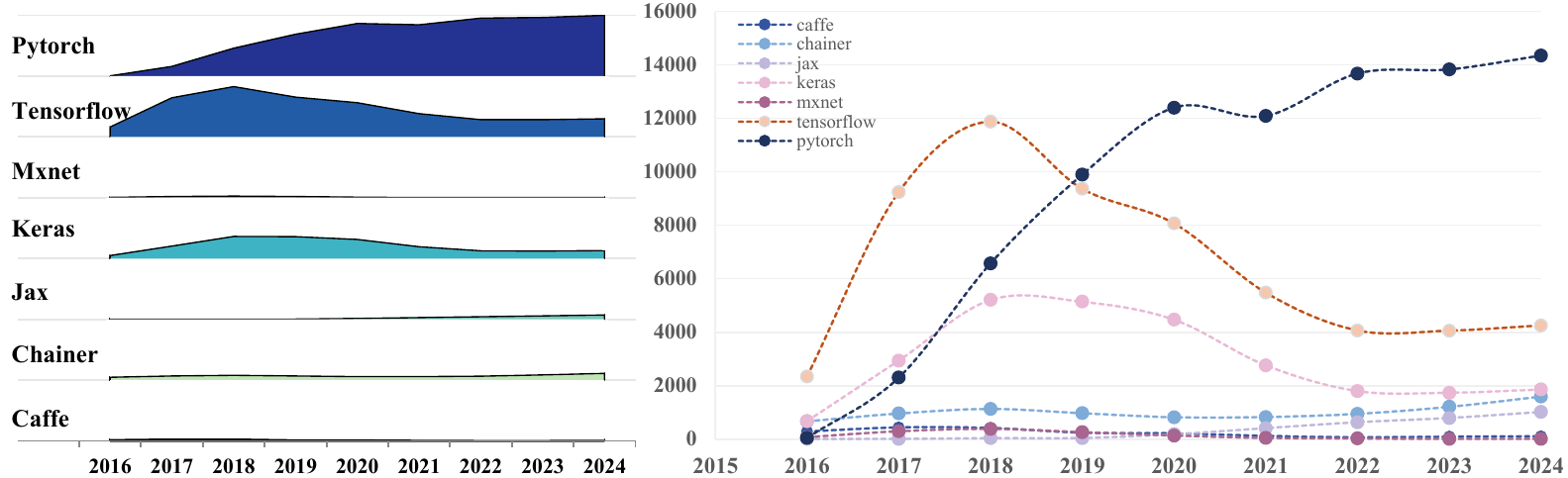}
    \caption{Trends of deep learning framework usage over years: (Right) absolute repository counts; (Left) normalized proportions across frameworks.}
    \label{fig: proportion}
    \vspace{-4mm}
\end{figure}

\subsection{Data Collection}
\label{d2}

\noindent \textbf{Repository Collection Scope and Rationale.}
Following prior mining software repositories works~\cite{ray2014large,borges2016understanding,munaiah2017curating,wu2023ossfp}, we collected open-source repositories from GitHub, the most popular platform for hosting open-source projects.  
A critical decision in defining our scope was the choice of the NN framework. The adoption of PyTorch has surged, with its usage in research papers growing from 29\% in 2019 to 65\% in 2023~\cite{hao2023torchbench}, and it now underpins over 70\% of AI research implementations~\cite{pytorch2024review}. As illustrated in Figure~\ref{fig: proportion}, which shows the trends of deep learning framework usage, the number of PyTorch-related repositories has consistently grown since 2018, while repositories for other frameworks have generally seen a decline or stagnation.
Given PyTorch's clear dominance and to ensure a focused analysis by avoiding framework-specific inconsistencies that could introduce noise into evolutionary patterns, we concentrated our collection on PyTorch repositories. This approach allows for a deep dive into the most active and representative segment of the current NN open-source ecosystem.  
Our data collection process follows the method outlined in OSSFP~\cite{wu2023ossfp}.
Specifically, we use the official GitHub API to retrieve all repositories tagged with Python and containing the keyword "pytorch" in their topics or repository names. This initial query yielded 78,243 repository links with their associated metadata, such as topics, creation date, and fork count.

\noindent \textbf{Filtering rules.}
To curate a dataset representative of substantive NN software evolution, we applied a multi-stage filtering process:
\begin{enumerate}[leftmargin=*,topsep=0pt,itemsep=0ex]
    \item \textbf{Exclusion of Tutorial Projects:} Repositories with names or descriptions contraining keywords like ``tutorial'', ``example'', or ``demo'' were excluded. These often contain simplified, non-production code unsuited for studying real-world software evolution.
    \item \textbf{Removal of Trivial Repositories:} Projects lacking any identifiable NN modular code components (as defined in Section~\ref{d3}) were discarded, as they do not contribute to the component-level analysis of NNBOM.
    \item \textbf{Quality and Activity Filtering:} We employed a standard repository ranking tool~\cite{criticality_score}, to assess project activity and maintenance. The bottom 5\% of repositories based on this score\footnote{We chose 5\% as a practical trade-off, which should be sufficient to eliminate low-quality repositories without significantly affecting the validity of large-scale evolution analysis} were filtered out to minimize the impact of inactive or poorly maintained projects.
\end{enumerate}
After applying these filters, our dataset comprised 55,997 PyTorch repositories, forming the basis for NNBOM construction.

\noindent \textbf{Version collection.}
For each selected repository, we collected version information. Since most GitHub repositories mark releases or significant milestones using Git tags, we extracted these tags as distinct versions. For repositories without any Git tags, we designate their current main branch as a single representative version. All versions were downloaded via the GitHub API, resulting in a total of 93,647 distinct repository versions for analysis. Data collection scripts and additional repository details are available at~\cite{nnbom}.

\begin{figure} [t]
    \centering
    \includegraphics[width=\linewidth]{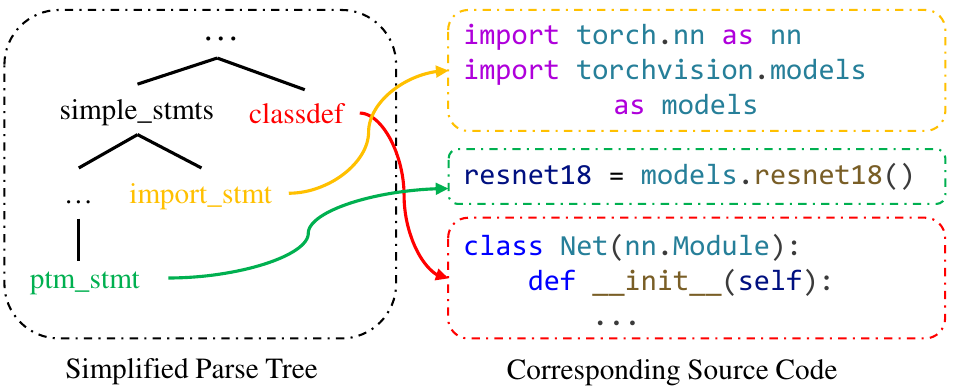}
    \caption{Extract specific components from the parse tree.}
    
    \label{fig: parsetree}
\end{figure}

\subsection{Component Analysis}
\label{d3}

The NNBOM for each repository is constructed by identifying and extracting three primary categories of NN software components: TPLs, PTMs, and neural network modules. Figure~\ref{fig: parsetree} illustrates how these components can be identified from source code, often via specific nodes in a parsed Abstract Syntax Tree (AST). The subsequent subsections detail the extraction process for each component type.

\subsubsection{TPL Extraction.}

TPLs are pre-packaged Python libraries, typically distributed via PyPI or WPILib, that provide essential functionalities. We extract TPL information using the two approaches below:

\begin{itemize}[leftmargin=*,topsep=0pt,itemsep=0ex]
\item \textbf{Extraction from Configuration Files:} We parse project configuration files (e.g., ``setup.py'', ``requirements.txt'') which often declare TPL dependencies and their versions. However, as noted by Peng~\et~\cite{peng2024less}, these files can be incomplete.

\item \textbf{Scanning Import Statements:} We use ANTLR to parse Python source code and identify all \codeff{import} statements. This captures TPLs used directly in code, though \codeff{import} statements typically do not specify version information.
\end{itemize}
To mitigate the limitations of each method, we combine these methods. Information from configuration files (including versions) is prioritized. The \codeff{import} statements are analyzed to distinguish between local project modules (e.g., relative imports like \codeff{import .module} or import matching project subdirectories) and external TPLs. The identified external TPLs from \codeff{import} statements are then merged with those from configuration files to create a comprehensive list for each repository version. This combined approach provides a more accurate depiction of TPL usage.


\subsubsection{PTM Extraction.}
\label{sec: invokedNNModels}

To save computational resources and leverage existing models, developers often reuse PTMs. We detect PTM invocations, particularly those from model hubs, by:
\begin{itemize}[leftmargin=*,topsep=0pt,itemsep=0ex]
    \item \textbf{Identifying Model Hub APIs:} Following Jiang~\et~\cite{jiang2022PTMempirical}, we target PTM invocations from eight popular model hubs: Hugging Face\cite{hugging-face}, TensorFlow Hub~\cite{tensorflowhub}, Model Zoo~\cite{modelzoo}, PyTorch Hub~\cite{PyTorch-Hub}, ONNX Model Zoo~\cite{ONNX-Model-Zoo}, Modelhub~\cite{Modelhub}, NVIDIA NGC~\cite{NVIDIA}, MATLAB Model Hub~\cite{matlab}. Six of these hubs provide APIs for PTM invocation; the other two (ModelZoo, ONNX Model Zoo) involve repository cloning, detected by comparing them against a database.
    \item \textbf{Recognizing LLM Deployment Frameworks:} We also identify PTM usage via large language model (LLM) deployment frameworks like vLLM~\cite{vllm}, DeepSpeed-MII~\cite{DeepSpeed-MII}, and CTranslate2~\cite{CTranslate2}. An example of such an invocation is shown in Figure~\ref{fig: TerminologyExample} (item \ding{175}).
    \item \textbf{Customized Parsing:} We modified Python's lexical (``lexer.g4'') and grammar (``parser.g4'') files to define rules specific to each hub's and framework's PTM invocation patterns. These customized grammars are used to build ASTs. During tree traversal, a symbol table tracks variable assignments. When a PTM invocation is identified, its path parameter is resolved (either directly if a string or via the symbol table if a variable) to capture the PTM source.
\end{itemize}
This systematic approach yields a detailed list of PTM invocations within the analyzed projects.


\subsubsection{Module Extraction}


In the PyTorch framework, an NN module is any class that inherits, directly or indirectly, from ``torch.nn.Module''. These are the fundamental building blocks of NN architectures. We extract these modules as follows:

When starting with a project, the initial step involves creating a symbol table and adding ``torch.nn.Module'' as an entry. Following this, we generate a parser tree for each Python file to aid in identifying classes and mapping their inheritance relationships. As we navigate through the classes in the project, any class whose superclass is already listed in the symbol table is also added to the table. This process is repeated until there are no further changes to the size of the symbol table, indicating that all relevant classes have been captured.
The classes captured in the symbol table, other than ``torch.nn.Module'', are recognized as the NN modules of the project.
To improve parsing efficiency, particularly when dealing with multiple versions of the same repository that contain numerous identical files, we implement an incremental parsing approach. Only files that differ between versions are parsed for each new version. The differences between consecutive versions are determined using the ``git log" command, which allows for faster processing while maintaining accuracy.

\subsubsection{Effectiveness of Component Extraction}\label{validation}
To evaluate the effectiveness of our component extraction pipeline, we conducted a manual validation study on a random sample of 100 repositories, which aligns with prior work~\cite{lambaria2022data,idowu2024large,sulun2023empirical}. Typically, about 100 repositories are randomly selected for manual inspection to balance representativeness and manageable workload. Specifically, we randomly selected 100 repositories stratified into five bins based on the total number of extracted modules, with 20 repositories sampled from each bin. This stratified sampling ensures coverage across repositories of varying complexity.
By comparing the manually annotated results with those generated by our method, we found that the identification accuracy for third-party libraries (TPLs) and neural network modules both reached 100\%, correctly identifying 2,431 TPLs and 1,271 neural network modules. The identification accuracy for pretrained models (PTMs) was slightly lower at approximately 98\%, with only 2 out of 83 instances missed due to dynamic specification of model information via console input, which is inherently unobservable through static analysis. Overall, the extraction method achieved a success rate of approximately 99.9\%, providing a strong foundation for the subsequent evolutionary analysis.
Since the primary objective of this study is to empirically investigate the neural network software ecosystem, detailed evaluation results are omitted. The complete source code and the manually verified dataset are available at~\cite{nnbom}.

\subsection{Data Processing}
\label{d4}
Once the raw component information (TPLs, PTMs, Modules) is extracted for all repository versions, further processing is necessary to construct the final NNBOM database and prepare it for evolutionary analysis. This involves module normalization and clone detection for dependency construction, index building for efficient data retrieval, and feature generation for both versions and modules.

\begin{table}[t]
    \centering
    \small

    \caption{Normalization rules}
    \label{tab: norm}

       \resizebox{0.5\textwidth}{!}{
       \begin{tabular}{c|c}
        \toprule
        \textbf{Rule} & \textbf{Description} \\ \midrule
        \textit{remove}  & Remove all comments and whitespace, including blank lines \\
        \textit{rename}  & Standardize variable names to a consistent naming convention \\
        \textit{replace} & Replace numeric and string literals with specified placeholders \\ \bottomrule
        \end{tabular}
        }
\end{table}

\subsubsection{Module-based Dependency Construction}\label{dependency_graph}
To understand component reuse and infer inter-repository dependencies, we identify cloned NN modules. A module in repository A is considered a reuse of a module from repository B if it is detected as a clone and appears later in time, following the timestamp-based inference standard in prior studies~\cite{zhang2013towards, jahanshahi2024beyond}. Our study focuses on detecting Type-1 (exact textual clones) and Type-2 (syntactic clones), as these forms of reuse are both prevalent and reliable. Empirical evidence confirms that Type-1/2 clones constitute the dominant types of reuse in open-source software~\cite{feng2024machine}, while Type-3 and Type-4 clones have less consistent definitions and less reliable detection, posing a significant risk of introducing noise and false positives in large-scale analysis~\cite{roy2018benchmarks,yu2019neural}. By limiting our analysis to high-confidence Type-1/2 clones, we ensure that the findings on module reuse are both scalable and reliable.
For clone detection, we adopt well-established tools~\cite{ML_is_you_need, tacc, sajnani2016sourcerercc, nicad}, applying NN module-specific normalization. These methods have demonstrated high accuracy in detecting Type-1/2 clones, thereby supporting the robustness of our database construction and subsequent empirical analysis.
The normalization rules, detailed in Table~\ref{tab: norm}, are applied to each extracted NN module before hashing. These steps reduce superficial differences, allowing for more accurate identification of functionally similar or identical modules. After normalization, a hash value is generated for each module. Modules sharing the same hash are grouped into a ``clone family''. If two different repositories contain modules belonging to the same clone family, we consider a dependency link to exist between them, mediated by the shared module. Note that this mechanism establishes dependencies between repositories, not between different versions of the same repository.

\subsubsection{Index Building}
In this phase, we aim to establish indices for both modules and versions. By creating these indices, we can quickly retrieve various types of information about the version corresponding to each module. This enables us to efficiently analyze clone families, determining their application domains (by aggregating the domains of all modules), their lifespan (the difference between the earliest and latest module creation years), and more.
Additionally, these indices allow us to swiftly obtain clone information between versions, facilitating the construction of the dependency graph.

\subsubsection{Feature Generation}\label{keyword}
Finally, we generate a structured set of features for each version and each unique NN module in the NNBOM.

The version features are as follows: 
\begin{itemize}[leftmargin=*,topsep=0pt,itemsep=0ex]
\item \textbf{Version Index.} The index of the version.
\item \textbf{TPL List.} The TPLs that the version depends on. The TPLs extracted from configuration files contain version information, whereas those extracted from \codeff{import} statements do not.
\item \textbf{PTM List.} The PTMs invoked by the version. 
\item \textbf{Self-Developed Module List.} The index of the modules developed independently within the version.
\item \textbf{Cloned Module List.} The index of the cloned modules within the version.
\item \textbf{Release Time.} The release time of the version. This feature is vital for determining the origin of the module and conducting an evolutionary analysis.
\end{itemize}

The module features are as follows: 
\begin{itemize}[leftmargin=*,topsep=0pt,itemsep=0ex]
\item \textbf{Module Index.} The index of the module. 
\item \textbf{Module Hash.} The hash value of the normalized module. It ensures accurate mapping of module pairs without structural and semantic differences. 
\item \textbf{Version Index.} The version from which the module originates. 
\item \textbf{Lines of Code (LoC).} The number of code lines in the module, including blank lines and comments.
\item \textbf{Domains.} 
To analyze the application scope of the AI software module, repositories are classified into seven representative domains: Unsupervised Learning (UL), Reinforcement Learning (RL), Computer Vision (CV), Multimodal Learning (MML), Natural Language Processing (NLP), Generative Models (GM), and Transformer (Trans). These domains were selected by jointly considering top categories from the Model Zoo platform~\cite{modelzoo} and prevalent AI research areas. For accurate classification, a list of representative keywords for each domain was manually curated with domain expert assistance (detailed in our replication package~\cite{nnbom}). A repository is assigned to a domain if its name or  GitHub topics match any of these keywords. Repositories can be assigned to multiple domains if they match keywords from several categories, or to no domain if relevant metadata is insufficient or missing. Modules inherit the domain(s) of their parent repository.
\item \textbf{Module Frequency.} The number of times the module has been cloned in the GitHub community. 
A higher number of clones indicates a greater popularity of the module.
\end{itemize}

\subsubsection{Overview of the Constructed \texorpdfstring{\textbf{\underline{NNBOM Dataset}}}{NNBOM Dataset}}
Upon completion of the data collection, component analysis, and processing pipeline, the final NNBOM dataset is compiled. This dataset encompasses detailed component information for each repository version and an inter-repository dependency graph. In this graph, nodes represent individual repositories, and weighted edges signify the number of shared module clone families between them, derived as described in Section~\ref{dependency_graph}. The resulting NNBOM dataset provides a comprehensive view of component-level composition and inter-repository dependencies within the studied PyTorch ecosystem. It comprises data from 93,647 repository versions originating from 55,997 curated repositories. This dataset includes a total of 1,894,763 TPLs, 23,559 PTM invocations, and 3,102,311 NN modules. Across these repositories, we identified 2,135,351 version dependency edges based on the shared module clones, highlighting extensive component reuse and cross-project propagation. This rich NNBOM dataset serves as the foundation for the empirical evolution analysis presented in the subsequent sections.

%% file: chapters/RQ1.tex
\begin{table}
\centering
\caption{Annual  changes in overall trends}
\label{tab:general_trend}
\begin{tabular}{c|cccc|cc}
\hline
\multirow{2}{*}{\textbf{Year}} & \multicolumn{4}{c|}{\textbf{Total}} & \multicolumn{2}{c}{\textbf{Average}} \\
                               & Version  & TPL     & PTM   & Module & Module            & LoC              \\ \hline
\textbf{2017}                  & 1,414     & 49,045   & 131   & 20,454  & 14.47             & 54.09            \\
\textbf{2018}                  & 5,089     & 138,535  & 981   & 42,059  & 8.26              & 49.95            \\
\textbf{2019}                  & 8,737     & 222,860  & 1,031  & 107,275 & 12.28             & 49.77            \\
\textbf{2020}                  & 12,833    & 291,988  & 2,154  & 202,197 & 15.76             & 50.57            \\
\textbf{2021}                  & 15,267    & 299,193  & 2,393  & 440,779 & 28.87             & 52.35            \\
\textbf{2022}                  & 16,468    & 317,097  & 3,641  & 512,857 & 31.14             & 55.75            \\
\textbf{2023}                  & 16,684    & 290,761  & 5,475  & 685,572 & 37.29             & 56.57            \\
\textbf{2024}                  & 17,155    & 285,284  & 7,753  & 925,296 & 62.58             & 46.81            \\ \hline
\end{tabular}

 \vspace{-4mm}
\end{table}

\section{RQ1:  Software Scale Evolution}
\label{sec:RQ1}


\subsection{Analysis Method}
To address \textbf{RQ1}, we first analyze the overall evolutionary trend of neural network software across dimensions: the total number of repository versions (\textit{Total Version}), the count of distinct third-party libraries (\textit{Total TPL}), the number of invoked pretrained models (\textit{Total PTM}), and the total count of modules (\textit{Total Module}).
Subsequently, we examine the scale characteristics of modules, the core NNBOM component, by analyzing the average number of modules per version (\textit{Average Module}) and the average lines of code per module (\textit{Average LoC}). Additionally, we investigate the distribution of version sizes, based on module counts, to capture structural variation across repositories.

\subsection{Results and Findings}


\noindent \textbf{Overall Trend.} Table~\ref{tab:general_trend} (Total column) presents the overall development trend within the PyTorch community. The first column reveals a consistent increase in neural network versions since PyTorch’s 2016 release, highlighted by a remarkable growth of 807.57\% from 2017 to 2020, signifying a period of rapid expansion.
The subsequent columns in Table~\ref{tab:general_trend} detail the evolution of each NNBOM component. Specifically, the number of distinct TPLs utilized per year increased substantially before 2020. After this period of rapid expansion, the annual growth in the number of TPLs used across the ecosystem significantly slowed, with overall usage patterns stabilizing, indicating a maturation of the TPL ecosystem. In contrast, the total number of PTM invocations and NN modules have both continued to grow steadily. Although overall PTM usage has increased, the number of direct PTM invocations remains relatively low; further analysis reveals that only 7.6\% of repositories include such invocations.
Considering the extensive presence of clone modules identified during our dataset construction, it is evident that \textit{researchers often prefer cloning models over directly invoking PTMs when developing neural network systems}. This preference likely reflects a desire for greater flexibility and control in model customization.
\begin{tcolorbox}[size=title]
	{ \textbf{Finding 1:}
    All NNBOM components exhibit continuous growth in scale. TPLs experienced rapid expansion until 2020, after which their usage stabilized, indicating a maturation of the TPL ecosystem. Meanwhile, PTMs and Modules show sustained growth.
}
\end{tcolorbox}

\begin{figure}
\centering
\includegraphics[width=\linewidth]{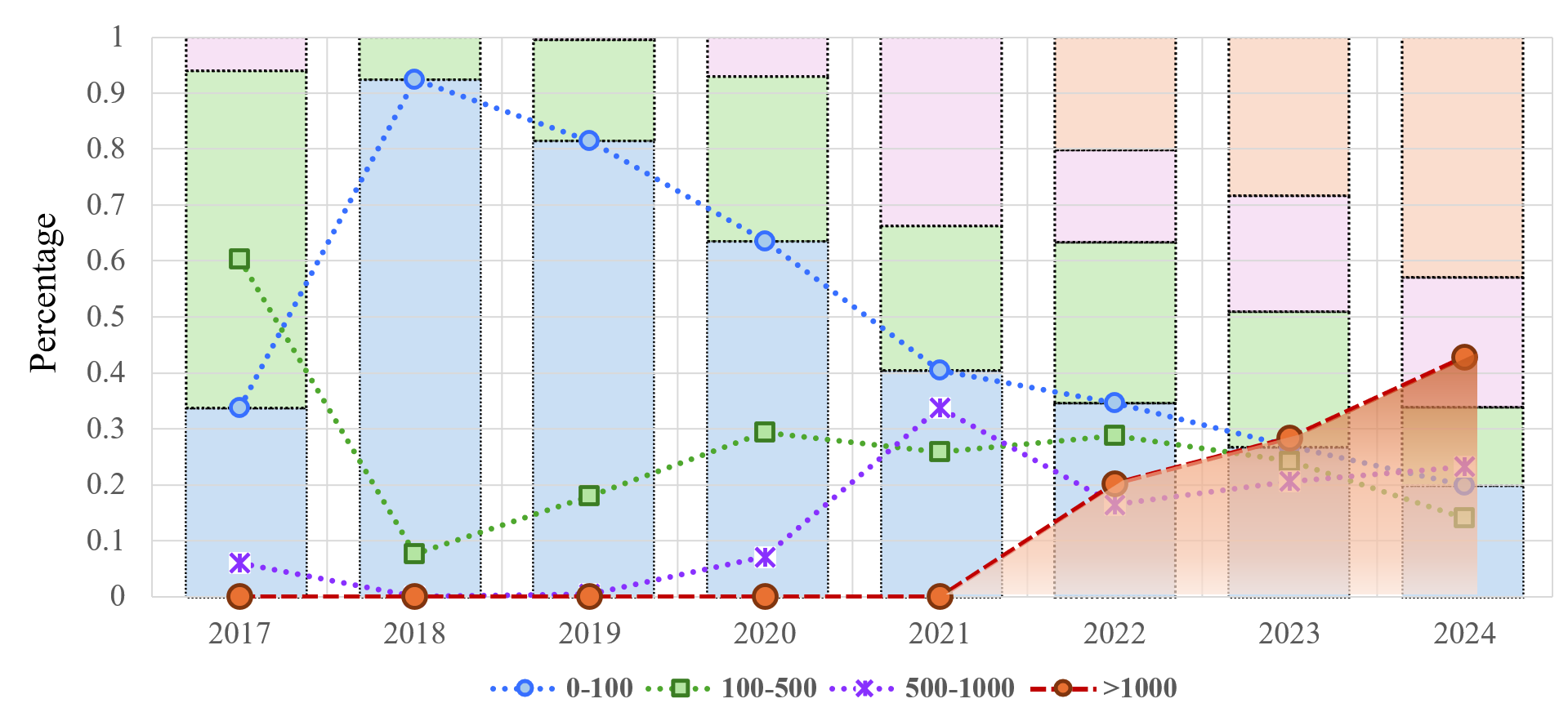}
\caption{Proportional changes in versions of different sizes.}
\label{fig: rq2-2}
\end{figure}

\noindent \textbf{Intra-Project Module Scale and Size.}
Since a module is essentially a self-contained code block, its total count is naturally influenced by the increasing number of repository versions. To decouple this effect and understand project-level characteristics, we compute the average number of modules per version and the average lines of code (LoC) per module, as shown in Table~\ref{tab:general_trend} (Average columns). The former metric reflects the module scale of each version, while the latter captures the average size of a module.
From 2018 onward (excluding 2017 due to early-stage data instability), the average number of modules per version has steadily increased, indicating a clear trend toward larger project sizes. Conversely, the average LoC per module has remained relatively stable throughout the study period, consistently hovering around 50 lines. This stability suggests that as neural network tasks grow in complexity, developers tend to decompose functionality into a greater number of smaller modules, rather than constructing larger, monolithic modules. The consistent module size of around 50 lines may reflect an implicit developer preference or convention for module granularity.
Figure~\ref{fig: rq2-2} provides a more detailed view, illustrating the annual distribution of repository versions categorized by their module count. Here, we categorize version sizes into four intervals: \textit{0--100}, \textit{100--500}, \textit{500--1000}, and $>$1000 modules. Except for the early-stage year 2017, a clear decline is observed in smaller versions (0--100) starting from 2018, suggesting the gradual maturation of projects beyond initial prototypes. Versions containing 100--500 and 500--1000 modules have become increasingly prevalent over time. Notably, from 2022 onward, there has been a sharp rise in the proportion of large-scale versions ($>$1000 modules). This surge coincides with the emergence of large language models and marks the onset of a new era in ultra-scale neural network development. This trend highlights \textit{a substantial shift toward building larger and more architecturally complex models}, driven by field advances and increased computational capabilities.

\begin{tcolorbox}[size=title]
	{ \textbf{Finding 2:}
The increasing average number of modules per version, coupled with a stable average module LoC (around 50 lines), indicates a consistent development convention: developers are addressing growing task complexity by decoupling projects into a greater number of smaller modules.
}
\end{tcolorbox}

%% file: chapters/RQ2.tex
\section{RQ2: Component Reuse Evolution}
\label{sec:RQ2}

\begin{figure}
\centering
\includegraphics[width=0.6\linewidth]{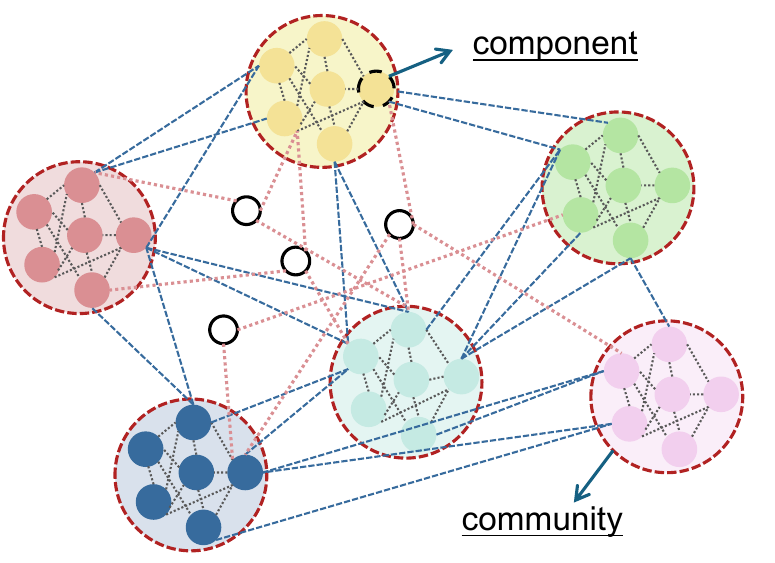}
\caption{Example of a component co-occurrence network.}
\label{fig:rq2-3}
\end{figure}

\subsection{Analysis Method}
To address RQ2, we design a two-fold exploratory analysis to examine the reuse evolution of neural network components.
First, as illustrated in Figure~\ref{fig:rq2-3}, we construct annual component co-occurrence networks. Each node in these networks represents a unique NNBOM component, and edges indicate strong co-usage, defined as co-occurrence in at least five different repositories. We then apply the Louvain algorithm~\cite{louvain} to partition these networks, maximizing modularity to identify communities. Each community represents a cluster of components frequently used in conjunction. By tracking the number and average size of these communities over time, we analyze the evolution of co-usage patterns and the structural organization of component dependencies.
Second, we focus on Modules, the core components of neural networks, and examine how the most frequently reused ones evolve over time. This analysis aims to uncover shifts in architectural preferences and highlight emerging functional trends in neural network design.

\subsection{Results and Findings}

\begin{figure}
\centering
\includegraphics[width=\linewidth]{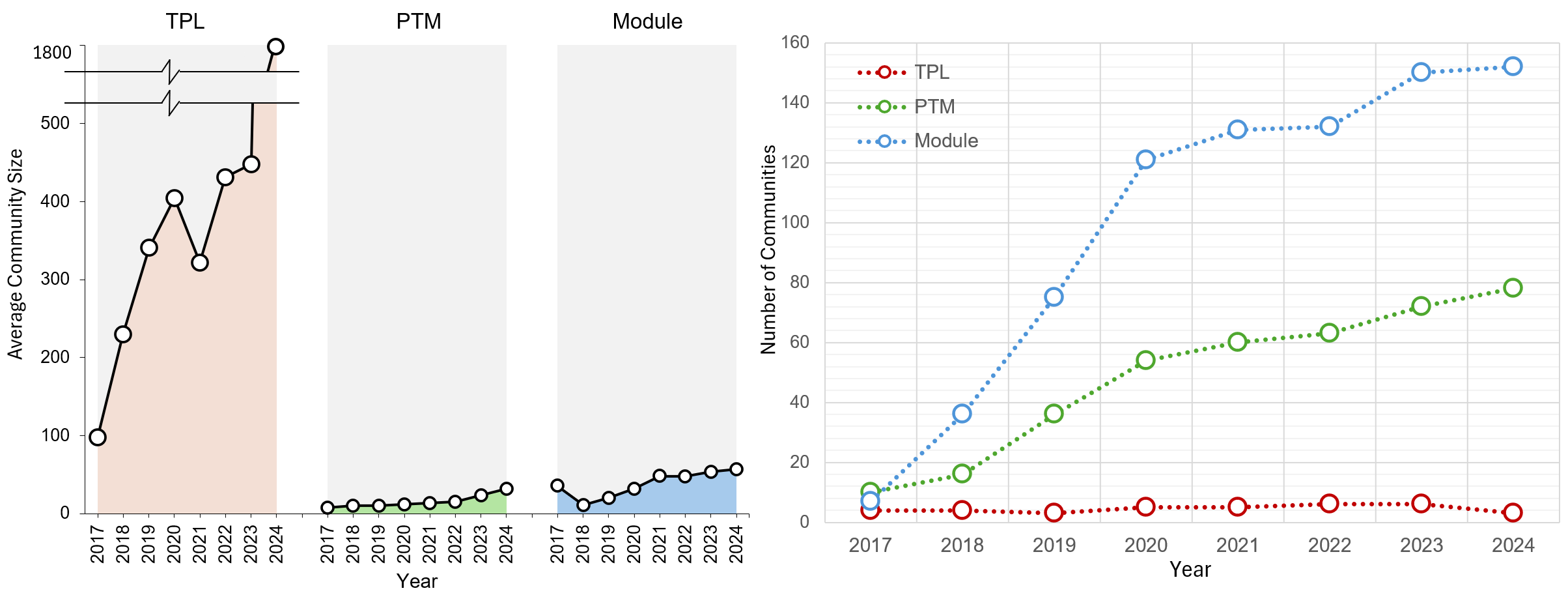}
\caption{Yearly trends of co-usage community count and average size, reflecting the evolving structure of component reuse.}
\label{fig:ocure}
\end{figure}

\noindent \textbf{Co-usage Community Dynamics.}
Figure~\ref{fig:ocure} illustrates the temporal trends in both the number and average size of co-usage communities for each component type.
First, the average community size exhibits a consistent upward trend across all component types, with TPLs exhibiting the fastest growth. This indicates increasing complexity in the co-occurrence networks, where components become more interconnected and form larger co-usage clusters. The pattern reflects a growing tendency for components, especially TPLs, to be reused alongside a broader set of counterparts, driven by accumulated developer experience and the gradual convergence of reusable modules into larger, and more integrated communities.
Second, the number of communities displays divergent trends. Both PTMs and Modules show sustained growth in community count, with Modules increasing most rapidly. This suggests that, alongside forming larger clusters, these components continue to diversify by supporting emerging functionalities and spawning new co-usage groups. In contrast, the number of TPL communities remains relatively stable over time. This stability, combined with growing community size, implies that TPLs are maturing into stable, general-purpose clusters with fewer novel configurations emerging.

\begin{tcolorbox}[size=title]
	{ \textbf{Finding 3:}
    All NNBOM components are increasingly integrated into larger co-usage communities, while PTMs and Modules additionally drive the emergence of new functional clusters.
}
\end{tcolorbox}

\noindent \textbf{Functional Evolution of Modules. }
To examine the functional evolution of neural network Modules, we construct a temporal evolution tree based on the top ten most reused Modules for each year from 2017 to 2024. These Modules reflect the dominant architectural patterns of each era. Their transitions reveal three key evolutionary phases, as shown in Figure~\ref{fig: evaluation}.

\textit{Initial Phase (2017).}
This stage features frequent reuse of simple functional Modules such as \texttt{sigmoid} and \texttt{padding}, reflecting PyTorch’s early requirement to encapsulate even basic operations within \texttt{nn.Module}.

\textit{Growth Phase (2018–2020).}
CNN-based architectures, especially ResNet and Inception, become mainstream. Their components are widely reused, often through direct cloning rather than invoking pre-trained models, enabling structural customization.

\textit{Transformer Expansion Phase (2021–2024).}
This phase witnesses a surge in Transformer-based Modules and increasing structural diversity. Developers increasingly build upon existing Transformer infrastructures, exemplified by the rise of modules such as \texttt{OnlyMLMHead} and \texttt{Attention} blocks. These Transformer-centric modules mark the beginning of the large model era, underpinning the development of a new wave of commercial-scale pre-trained systems.

\begin{tcolorbox}[size=title]
	{ \textbf{Finding 4: }
Module reuse has evolved from simple functional operations to Transformer-centric architectures, reflecting a structural transition toward scalable designs that underpin the emergence of modern large-model systems.

 }
\end{tcolorbox}

%% file: chapters/RQ3.tex
\section{RQ3: Inter-Domain Dependency Evolution}
\label{sec:RQ3}

\begin{figure}
    \centering
    \includegraphics[width=\linewidth]{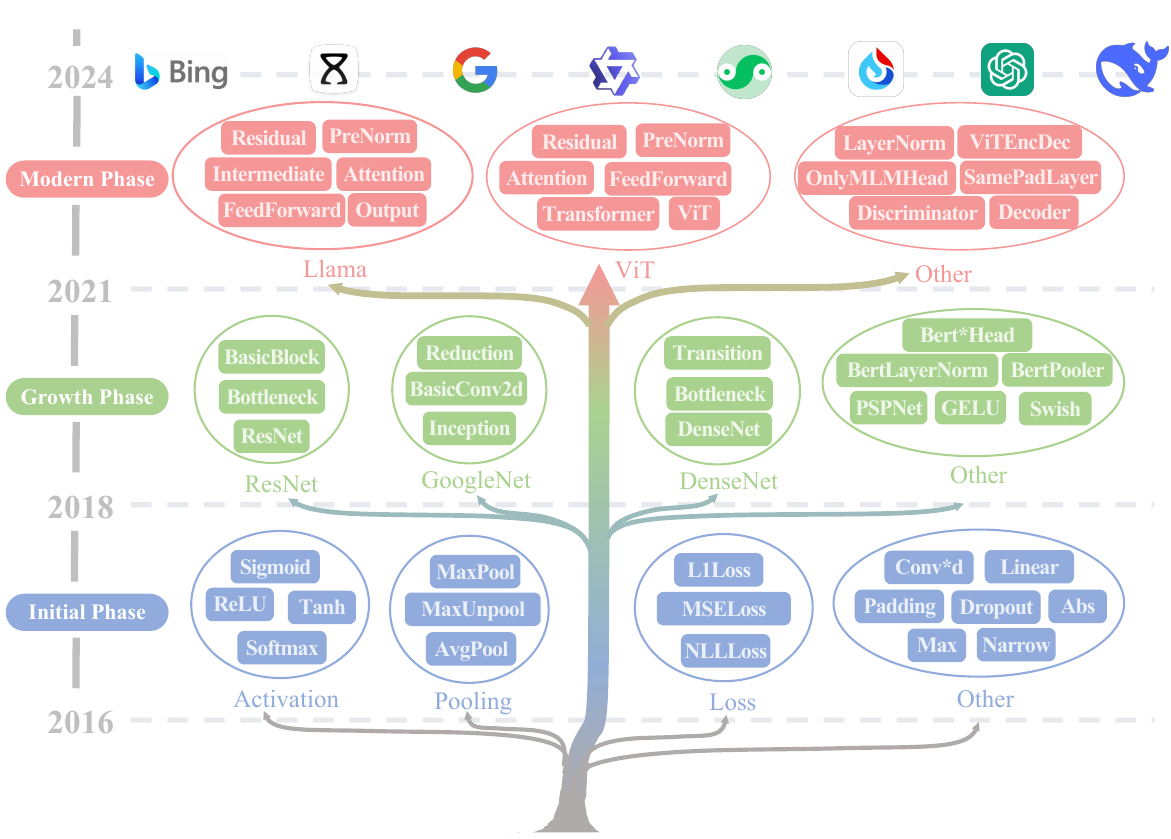}
    \caption{Evolution Tree: Tracing the Development of Neural Network.}
    \label{fig: evaluation}
\end{figure}

\subsection{Analysis Method}
To address \textbf{RQ3}, we conduct a three-part analysis focusing on: (1) annual changes in average entropy, (2) detailed trends in inter-domain dependency, and (3) the correlation between module lifespan and its domain usage range.

First, we calculate the annual average entropy to uncover the overall trend in inter-domain module dependency over time. The average entropy $\bar{H}$ for each year is computed as follows:
\begin{equation}
    \label{equ:entropy}
    \bar{H} = \frac{1}{N} \sum_{i=1}^N H_i, \quad H_i = -\sum_{k=1}^{K} p_k \log(p_k)
\end{equation}
where $N$ is the number of clone families in a given year, $H_i$ denotes the entropy of the $i$-th clone family, and $p_k$ is the proportion of modules in the $k$-th domain. 
A higher $\bar{H}$ indicates greater overall diversity in how module families span across different domains, suggesting increased inter-domain reuse.

After identifying the overall trend, we analyze specific inter-domain dependencies by examining module overlap among our seven representative domains. We identify the top five domain pairs with the highest degree of module sharing each year. The degree of overlap is quantified using:
\begin{equation}
    \label{equ:percentage}
    P_{AB} = \frac{|A\cap B|}{|A\cup B|}
\end{equation}
where $A$ and $B$ represent two domains, and $P_{AB}$ indicates the proportion of shared modules between them.

Finally, we investigate the correlation between a module's lifespan and its domain usage range. Here, a module’s lifespan is defined as the number of years between the earliest and latest versions in its clone family, while the domain range is measured by the number of distinct domains in which the family appears.

\subsection{Results and Findings}

\noindent\textbf{Annual Changes in Average Entropy. }
\begin{table}[t]
\caption{Annual changes in average entropy}
\label{tab: overrallEntropy}

\resizebox{\linewidth}{!}{
\begin{tabular}{ccccccccc}
\toprule
\textbf{Year} & 2017  & 2018  & 2019  & 2020  & 2021  & 2022  & 2023 & 2024  \\ \midrule
\textbf{Average Entropy}    & 0.157 & 0.175 & 0.231 & 0.265 & 0.298 & 0.385 & 0.457 & 0.485 \\ \bottomrule
\end{tabular}
}
\end{table}
Table~\ref{tab: overrallEntropy} presents the annual average entropy of domain diversity for module clone families. A higher entropy value signifies greater inter-domain module reuse. The data reveals a consistent upward trend in average entropy from 0.157 in 2017 to 0.485 in 2024, indicating increasingly frequent and diverse reuse of module clone families across different AI domains. This trend suggests an evolving design philosophy: modules embodying general-purpose, suitable for reuse across diverse domains, are becoming more prevalent compared to highly task-specific implementations. This shift points towards a growing emphasis on modularity and reusability driven by the inherent agnosticism of foundation NN components to specific tasks.

\begin{tcolorbox}[size=title]
	{ \textbf{Finding 5:}
Rising entropy suggests increasing cross-domain reuse and a shift toward general-purpose modules.
}
\end{tcolorbox}

\begin{table}[t]
\caption{Top 5 domain pairs by shared modules}
\label{tab:inter-domain}
\small
\centering
\resizebox{\linewidth}{!}{
\begin{tabular}{cccccc}
\toprule
\textbf{Year} & \textbf{Rank 1}    & \textbf{Rank 2}    & \textbf{Rank 3}    & \textbf{Rank 4}    & \textbf{Rank 5}    \\ \midrule
2017 & U-G-17.54 & C-G-12.14 & C-U-7.33  & N-G-5.98  & C-N-4.68  \\ 
2018 & C-G-14.02 & T-N-12.52 & G-M-5.03  & C-N-3.95  & N-G-2.84  \\ 
2019 & T-N-16.25 & U-G-11.07 & C-G-10.63 & N-R-6.37  & T-C-6.09  \\ 
2020 & T-N-21.31 & T-C-8.23  & C-G-8.17  & U-G-6.89  & C-N-6.23  \\ 
2021 & T-N-25.34 & T-C-12.13 & C-N-6.45  & C-G-5.37  & T-M-3.94  \\
2022 & T-N-27.87 & T-C-19.36 & C-N-8.79  & T-M-7.88  & N-M-4.61  \\ 
2023 & T-N-29.86 & T-C-19.94 & T-G-12.23 & T-M-10.08 & C-N-9.01  \\ 
2024 & T-N-29.56 & N-G-21.16 & T-G-17.25 & T-M-16.32 & T-C-10.68 \\ 
\bottomrule
\end{tabular}
}

\vspace{1mm}
\noindent
\parbox{\linewidth}{
\footnotesize
\textit{Note.} $\mathcal{D}_1$--$\mathcal{D}_2$--$\rho_{\mathcal{D}_1 \mathcal{D}_2}$ indicates that $\rho_{\mathcal{D}_1 \mathcal{D}_2}\%$ of modules are shared between domains $\mathcal{D}_1$ and $\mathcal{D}_2$.
}
\vspace{-3mm}
\end{table}

\noindent\textbf{Details of Inter-Domain Dependency. }
The shared module proportions across domain pairs offer empirical insight into the architectural evolution of neural network software. 
Table~\ref{tab:inter-domain} presents the top five domain pairs with the highest frequency of cross-referenced modules each year.
In 2017, the top-ranked pairs were UL-GM (17.54\%), CV-GM (12.14\%), and CV-UL (7.33\%), indicating that module reuse was initially concentrated in unsupervised learning, generative modeling, and computer vision.
The overlap among these three domains reflects a characteristic of the \textbf{early stage}: the dominance of GAN-based unsupervised image generation frameworks, primarily applied to visual tasks. 
Notably, NLP-related structures were absent from the top reuse pairs, as large-scale text generation had not yet emerged as a major trend.

From 2018 onward, Transformer modules (Trans) emerged and rapidly gained prominence. The T-N (Transformer + NLP) pair immediately entered the top five in 2018 with a 12.52\% share and ranked first every year thereafter, peaking at 29.86\% in 2023. This sharp rise highlights the central role of NLP in reusing Transformer-based architectures, in line with the emergence of attention-based language models. This marks the beginning of the \textbf{Transformer stage} in 2018.
Following this, the Transformer architectural paradigm began to diffuse into other domains. The diffusion trajectory unfolded in distinct phases:
The first major wave occurred in computer vision. The T-C (Transformer + CV) pair appeared in 2020 (8.23\%) and steadily increased to 19.94\% by 2023, driven by the adoption of Vision Transformers (ViT) and Transformer-based object detection models such as DETR. This trend reflects a structural migration in CV toward Transformer-based backbones.
A second wave of diffusion emerged in the domains of generative modeling and multimodal learning. The T-M (Transformer + Multimodal) pair rose from 7.88\% in 2022 to 16.32\% in 2024, while T-G (Transformer + GM) entered the top five in 2023 and reached 17.25\% in 2024. These shifts correspond to the rise of large-scale language models and a transition in generative modeling from image-centric synthesis to Transformer-based generation.
Most notably, by 2024, the N-G (NLP + GM) pair surpassed the traditional C-G (CV + GM), which dropped out of the top five. This indicates a new convergence: generative tasks are increasingly grounded in NLP-centric, Transformer-driven architectures, reflecting a realignment of generative AI around language-first paradigms.
Together, these trends depict a multi-phase diffusion trajectory of the Transformer paradigm: from its NLP origins to widespread structural integration across vision and generative tasks, \textit{culminating in the convergence of language and generation as the dominant axis in modern neural network development.}

\begin{tcolorbox}[size=title]
	{ \textbf{Finding 6:}
Cross-domain dependencies evolved from early GAN-based visual generation to a Transformer-driven phase marked by widespread architectural diffusion and a shift toward text generation.
}
\end{tcolorbox}
\begin{table}
\caption{Module count by lifespan and domain}
\label{tab: lifespan}
\small
\begin{threeparttable}
\resizebox{0.90\linewidth}{!}{
\begin{tabular}{c|ccccccc}
\toprule
\textbf{L.$\backslash$ D.} & 1     & 2     & 3    & 4    & 5   & 6  & 7 \\ \midrule
1 & 74373 & 21030 & 3553 & 673 & 54 & 2 & 0 \\
2 & 7515 & 5214 & 1833 & 343 & 41 & 3 & 0  \\
3 & 1863 & 2530 & 869 & 526 & 289 & 30 & 4\\
4 & 826 & 1016 & 483 & 399 & 52 & 14 & 2\\
5 & 397 & 547 & 254 & 189 & 84 & 25 & 0\\
6 & 134 & 157 & 172 & 51 & 40 & 18 & 5\\
7 & 17 & 53 & 66 & 27 & 17 & 40 & 8\\
8 & 0 & 3 & 16 & 4 & 6 & 4 & 2\\
\bottomrule
\end{tabular}
}
\begin{tablenotes}
\footnotesize
    \item L.: Lifespan (number of years); D.: Domain Range.
\end{tablenotes}
\end{threeparttable}
\vspace{-4mm}
\end{table}

\noindent\textbf{Correlation Between Module Lifespan and Domain Range.}
The third perspective shifts focus to the relationship between module lifespan and cross-domain usage. Table~\ref{tab: lifespan} presents the number of modules categorized by their lifespan (i.e., number of years observed) and domain range (i.e., number of distinct domains in which they are used).
The results reveal a clear positive correlation between lifespan and domain range: modules reused across more domains tend to have significantly longer lifespans. For example, when the domain range is limited to one, 74,373 modules have a lifespan of only one year, while none survive for eight years. This pattern demonstrates that most modules are both short-lived and domain-specific, with counts decreasing sharply as lifespan increases.
Conversely, fixing the lifespan to one year (i.e., the first row), the number of modules drops rapidly as domain range increases, further confirming that broadly reused modules are rarely short-lived. This bidirectional trend suggests that the ability to generalize across domains is a strong indicator of a module’s utility and longevity.
To illustrate this pattern, consider the most notable modules that exhibit both the longest lifespan of eight years and usage across up to seven domains. Although rare, these modules are highly representative. According to NNBOM, they include \texttt{Bottleneck} and \texttt{BasicBlock}, two foundational building blocks introduced in ResNet and widely reused as backbones across a variety of neural architectures. Their sustained presence across time and domains reflects both their functional robustness and broad architectural compatibility.



\begin{tcolorbox}[size=title]
	{ \textbf{Finding 7:} 
 A wider domain range indicates stronger functionality and broader applicability, often correlating with longer lifespans. Cross-domain adaptability thus serves as a promising and quantifiable indicator of module quality and long-term stability.
 }
\end{tcolorbox}

%% file: chapters/5.application.tex
\section{Application}
To demonstrate the practical utility of the NNBOM dataset and our analytical findings, we developed two prototype applications. These prototypes showcase how NNBOM can directly support software engineering tasks for both ecosystem maintainers and individual developers.

\noindent\textit{(1) Multi-repository Evolution Analyzer.}
This tool enables maintainers to dynamically assess the impact of newly added repositories on the broader neural network software ecosystem. It quantifies growth in TPLs, PTMs, and modules, as well as the expansion of module-level dependencies by distinguishing between reused and newly created modules. This offers a practical lens into the ongoing evolutionary dynamics of the ecosystem. 
\noindent\textit{(2) Single-repository Component Assessor and Recommender.}
Aimed at developers, this tool analyzes a target repository using NNBOM to assess its components, highlighting outdated or newly added modules. It also recommends potentially useful components based on co-usage patterns and identifies similar repositories based on overall component similarity.

\noindent \textbf{Case Study.} 
(1) For multi-repository analysis, we examine repositories added in December 2024 using the NNBOM constructed from data before November 2024. The tool identifies 94 previously unseen TPLs, 44 new PTMs, and 6,545 new Modules introduced by these repositories. Among a total of 14,467 Modules in this batch, 45.2\% are original creations, highlighting a notable level of innovation.

\noindent (2) For single-repository analysis, based on the API search for top-recommended repositories created after January 1, 2025, the repository \href{https://github.com/test-time-training/ttt-video-dit/tree/main}{\texttt{test-time-training/ttt-video-dit}} stands out. Created on April 13, 2025, it has quickly gained 1.6k stars and 133 forks. This repository contains 43 TPLs, no PTMs, and 40 modules, of which 32 are newly created and eight are reused. Among the reused modules, the oldest dates back to 2020, while 4 were introduced in 2024, indicating that reused modules typically originate from the preceding year.
Notably, it introduces a new TPL, \href{https://pypi.org/project/test-time-training/}{\texttt{test-time-training}}, released on February 11, 2025, which provides support for test-time training kernels. According to our similarity analysis based on module composition, the most similar repository is \href{https://github.com/Auto1111SDK/Auto1111SDK}{\texttt{Auto1111SDK/Auto1111SDK}}, suggesting potential architectural or functional overlap that developers can reference or further explore.

Detailed usage instructions and case studies are provided in~\cite{nnbom}.

%% file: chapters/threats_to_validity.tex
\section{Threats to Validity}
\subsection{Internal Validity}
\noindent\textbf{Domain Labeling Accuracy.}  
To assign domain labels for RQ3, we relied on keywords in repository names and topic tags. This heuristic, while capturing many cases, may overlook repositories with ambiguous or missing metadata. potentially introducing error. However, fully reliable automated domain inference from content is challenging and remains unavailable. Our approach, prioritizing precision via manually curated keywords (detailed in Section~\ref{keyword} and~\cite{nnbom}), minimizes noise in the inter-domain analysis.

\noindent\textbf{Toolchain Reliability.}  
Another potential threat concerns the effectiveness of the toolchain used during dataset construction. Metadata collection is straightforward. Module dependencies for RQ3 rely on Type-1/2 clone detection, previously validated for high accuracy~\cite{ML_is_you_need, tacc, sajnani2016sourcerercc,nicad} and applied with carefully designed normalization (Table~\ref{tab: norm}) to ensure precision. The primary potential bias is in component extraction (TPLs, TPMs, Modules). Our manual validation (Section~\ref{validation}) on a stratified sample of 100 repositories showed high overall accuracy ($\thicksim$99.9\%), with minor PTM identification errors (2 out of 83 due to dynamic runtime specification, a static analysis limitation) having negligible impact on dataset quality. For untagged repositories, using the main branch as a single version is a simplification that might underrepresent their evolution for RQ1, but ensures their inclusion; the majority of our versions derive from tags, mitigating this.

\subsection{External Validity}
Our analysis focuses exclusively on PyTorch repositories, which may not fully represent the entire NN software ecosystem. 
Other frameworks were excluded due to significant methodological challenges in cross-framework module normalization and PTM identification. Components often lack direct, reliable counterparts across frameworks, making unified analysis beyond our current scope. However, as discussed in Section~\ref{d2}, PyTorch is the dominant framework in AI research and open-source development. Thus, our focus on  55,997 curated PyTorch repositories, selected via rigorous filtering to include substantive projects, provides a strong and representative basis for analyzing contemporary NN software evolution, minimizing generalization bias within this leading ecosystem. 

%% file: chapters/8.conclusion.tex
\section{Conclusion}

In this study, we construct NNBOM, a large-scale dataset capturing the component composition and dependencies of 55,997 neural network repositories. Based on this foundation, we conduct a comprehensive empirical analysis of neural network software evolution across three dimensions: software scale, component reuse, and inter-domain dependencies. Our findings offer valuable insights into the structural and functional trends shaping open-source neural network development.
